\documentclass[12pt]{article}

\usepackage{url}
\usepackage{amsmath}
\usepackage{amssymb}
\usepackage[dvips]{graphicx}
\usepackage{rotating}
\usepackage{fancyhdr}

\def\lfig#1#2#3#4{
\begin{figure}
\centerline{\hfill \includegraphics[height=#3]{#2}\hfill}
\vspace{-0.5cm}
\caption{#1 \label{#4}}
\end{figure}
}

\newcounter{tabl}
\newcommand{\be}{\begin{equation}}
\newcommand{\ee}{\end{equation}}
\newcommand{\beq}{\begin{eqnarray}}
\newcommand{\eeq}{\end{eqnarray}}
\newcommand{\bea}[2]{\be\label{#2}\begin{array}{#1}}
\newcommand{\eea}{\end{array}\ee}

\newcommand{\bR}{\mathbb{R}}

\newcommand{\Nint}{\mathbb{N}}

\newcommand{\IC}{\mathbb{C}}

\newcommand{\Cmat}{{\mathbb C}}

\newcommand{\Umat}{{\mathbb U}}

\def\det{\,{\rm det}\, }

\def\tr{\,{\rm tr}\, }

\def\({\left(}
\def\){\right)}
\def\[{\left[}
\def\]{\right]}
\def\p{\partial}

\def\11{1\!\! 1}

\def\hf{\frac{1}{2}}
\def\hft{{\textstyle\frac{1}{2}}}

\def\eps{\varepsilon}

   \def\CI {{\cal I}}

   \def\CO {{\cal O}}

   \def\CU {{\cal U}}
   \def\CV {{\cal V}}

\newcommand{\im}{\gamma}

\newcommand{\SSA}{{\bf A}}

\newcommand{\Ppr}[2]{\pi^{(#1)}_{#2}}
\newcommand{\pr}[1]{\pi^{(#1)}}

\def\jp{j^+}
\def\jm{j^-}

\begin{document}
%
%
%
\title{The new vertices and canonical quantization}
\vspace{0.7cm}
\author{Sergei Alexandrov}
\date{}
\maketitle
\vspace{-1cm}
\begin{center}
\emph{Laboratoire de Physique Th\'eorique \&
Astroparticules, CNRS UMR 5207,}\\
\emph{Universit\'e Montpellier II, 34095 Montpellier Cedex 05, France}
\end{center}
\vspace{0.1cm}
\begin{abstract}
We present two results on the recently proposed new spin foam models.
First, we show how a (slightly modified) restriction on representations in the EPRL model
leads to the appearance of the Ashtekar-Barbero connection, thus bringing this model even closer to LQG.
Second, we however argue that the quantization procedure used to derive the new models
is inconsistent since it relies on the symplectic structure of the unconstrained BF theory.

\end{abstract}
%

\section{Introduction}

Path integral and canonical quantizations are two alternative ways to arrive at
quantum theory. They are known to be closely related to each other, each with
its own perquisites and disadvantages. In the context of background independent quantum gravity
they are represented by the spin foam (SF) approach \cite{Baez:1997zt,Perez:2003vx}
and loop quantum gravity (LQG) \cite{Thiemann:2001yy,Rovelli:2004tv}.
Although a qualitative relation between these two approaches has been understood long ago \cite{Reisenberger:1996pu},
at the quantitative level there was a striking disagreement.

The situation has improved thanks to the appearance of the new SF models \cite{Freidel:2007py,Engle:2007wy},
which replaced the Barrett-Crane model \cite{Barrett:1997gw} that was the leading proposal during ten years.
In particular, it was claimed that the state space of the EPRL model \cite{Engle:2007wy}
is identical to the kinematical Hilbert space of LQG, namely that the spin foam boundary states resulting from
imposing constraints of general relativity are given by $SU(2)$ spin networks.
This claim was however based just on a formal coincidence of the set of labels coloring
the states in the spin foam model and LQG. It was also supported by comparison of the spectra
of geometric operators, area and volume \cite{Ding:2009jq}. But the quantum operators
in the EPRL model are not uniquely defined and this ambiguity was essentially used
to make the spectra to reproduce the LQG results.

On the other hand, the states in both approaches can be seen as functionals of a connection variable
and, if the state spaces are the same, the functionals representing the same state must also be identical.
Such identification was missing so far. The situation is complicated by the fact that
the spin foam quantization is performed in terms of the spin connection $\omega^{IJ}$, whereas the LQG states
are constructed using the so called Ashtekar-Barbero (AB) connection $A^a$ \cite{Ashtekar:1986yd,Barbero:1994ap},
requiring moreover the imposition of a partial gauge fixing. As a result, the actual relation between
the states of the EPRL model and LQG is not so evident.

One of the goals of this paper is to elucidate this issue. To this end, we remark
that the spin connection projected to the subspace defined by the EPRL intertwiner
with a slightly modified restriction on representations naturally gives rise to the AB connection.
This follows from the simple fact
\be
\pr{j}K_a^{(\lambda)} \pr{j}=\beta_{\lambda,j}L_a^{(j)} ,
\label{relgen}
\ee
where $K_a^{(\lambda)}$ are boost generators in representation $\lambda$,
$\pr{j}$ is the projector on representation $j$ of the $SU(2)$ subgroup, $L_a^{(j)}$ are rotation generators
in this representation, and $\beta_{\lambda,j}$ is a number
depending on both $\lambda$ and $j$. The EPRL constraints ensure that $\beta_{\lambda,j}\approx \gamma$, {\it i.e.},
for large spins the proportionality coefficient approaches the Immirzi parameter.
We propose a simple refinement of the constraints, which amounts to choosing a different
ordering for Casimir operators thereby fixing this ambiguity, so that $\beta_{\lambda,j}= \gamma$. Then one has
\be
\pr{j}\(\omega_i^{IJ} J_{IJ}^{(\lambda)}\) \pr{j}=A_i^a L_a^{(j)},
\label{projcon}
\ee
where $J_{IJ}$ is the full set of generators of the gauge group. Thus, it is possible to recover
the LQG connection variable and this works for both, Lorentzian and Riemannian signatures.

Unfortunately, this is not the end of the story because we actually do {\it not} have the
combination \eqref{projcon} in the EPRL model. In this model the states are projected spin networks
\cite{Livine:2002ak,Alexandrov:2002br} which are constructed from the projected holonomies of the spin-connection
\be
\CU_\alpha^{(\lambda,j)}=\pr{j}U_\alpha^{(\lambda)}[\omega] \pr{j},
\qquad
U_\alpha[\omega]={\cal P}
\exp\left(\int_\alpha \omega^{IJ} J_{IJ}\right).
\label{projW}
\ee
Thus, to get the AB connection as in \eqref{projcon}, one needs to bring the projectors up to the exponential.
This can be achieved by their insertion into every point of the integration path which gives rise
to {\it fully projected} holonomies introduced in \cite{Alexandrov:2002xc}.
It is not clear what this procedure means from the SF point of view, but this seems to be the only way
to get the LQG Hilbert space from the EPRL model.

The second goal of this work is to reconsider the imposition of the simplicity
constraints in the new SF models. This is the crucial step which is supposed to turn
a SF model of BF theory into a theory of quantum gravity. It is usually done by promoting
the $B$ field of BF theory to a quantum operator identified with the generators of the gauge algebra
and then imposing the resulting quantum constraints on the state space of the BF SF model.

However, this procedure disagrees with the Dirac rules of quantization of constraint systems.
The reason is that, following this strategy, one quantizes
the symplectic structure of BF theory which is not the same as the symplectic structure
of general relativity. The problem occurs already at the level of imposing the diagonal simplicity constraint
so that the improved treatment of the cross simplicity in the new models does not solve this problem.
We illustrate some unphysical features resulting from the above mentioned strategy
on a simple example encoding the basic kinematics of general relativity.
As a result, we arrive at the disappointing conclusion that despite promising results both models
\cite{Freidel:2007py,Engle:2007wy} are quantum mechanically inconsistent,
with the only exception of the FK model {\it without} the Immirzi parameter by reasons to be explained below.

The organization of the paper is as follows.
In the next section we show how the EPRL restriction on representations gives rise to the AB connection.
We perform the analysis both in the Lorentzian and in the Riemannian cases.
In subsection \ref{subsec_proj} we present the construction necessary to get the projection leading to
the LQG states. Moreover, we generalize it to avoid the partial gauge fixing with important consequences
for the closure constraint in SF models.
Section \ref{sec_simple} is devoted to the analysis of the constraint imposition in the SF models.
We start with a very simple model demonstrating all characteristic features of the SF approach and
discuss its implications in subsection \ref{subsec_SF}.
The main conclusions can be found in section \ref{sec_concl}.

\section{EPRL vs. LQG}
\label{sec_connect}

\subsection{EPRL constraints and AB connection}

\lfig{Projected spin network and the structure of its intertwiners.}{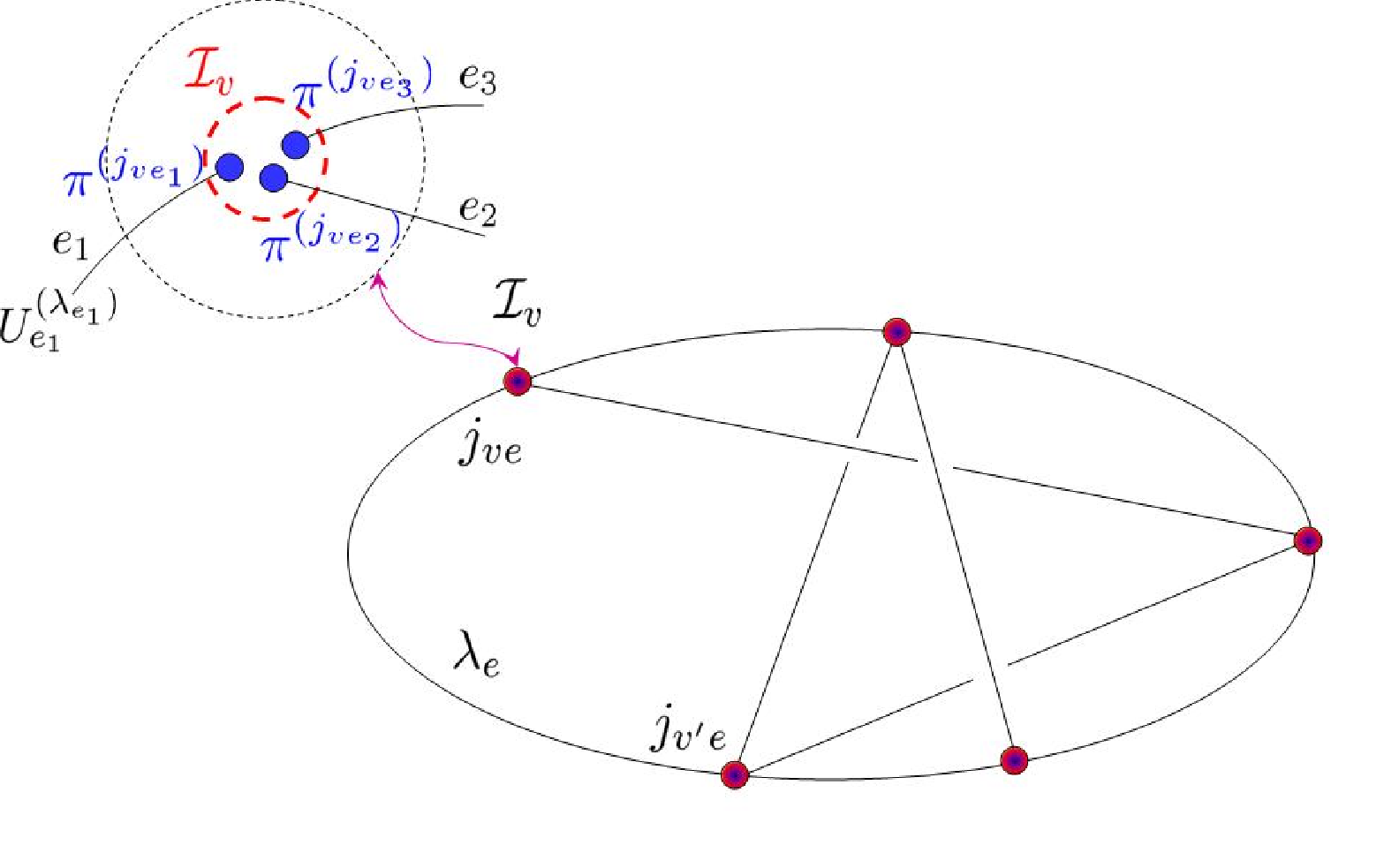}{6.4cm}{figprojsp}

The states of the EPRL model are described by a subset of projected spin network states.
In general, a projected spin network can be viewed as a graph
with the following coloring (see Fig. \ref{figprojsp}):
\begin{itemize}
\item edges carry representations $\lambda_e$ of the gauge group $G$;
\item every end of an edge (or the pair $(ve)$ with the vertex $v\subset e$) gets representation $j_{ve}$
of a subgroup $H$ appearing in the decomposition of $\lambda_e$ on $H$;
\item vertices are colored by $H$-invariant intertwiners $\CI_v$ coupling $j_{ve}$.
\end{itemize}
In our case $G=SL(2,\IC)$ or $Spin(4)$ depending on the signature and $H=SU(2)$.
Besides, the representations $\lambda_e$ and $j_{ve}$ are not arbitrary but they are restricted to satisfy certain constraints
representing a quantized version of the simplicity constraints of Plebanski formulation of general relativity.
The constraints can be split into two classes: the diagonal and cross simplicity.
The former gives restrictions on representations $\lambda_e$, whereas the latter produces conditions on $j_{ve}$ \cite{Engle:2007wy}
\be
\begin{split}
& \(1+\frac{\sigma}{\im^2}\)C^{(2)}_G(\lambda_e)-\frac{2\sigma}{\im}\,C^{(1)}_G(\lambda_e)=0,
\\
&  C^{(2)}_G(\lambda_e)=2\gamma C_H(j_{ve}),
\end{split}
\label{Casconstr}
\ee
where $\sigma=\pm 1$ for Riemannian (resp. Lorentzian) signature and the Casimir operators are defined as
\be
C_G^{(1)}=\hf\, J \cdot J ,
\qquad
C_G^{(2)}=\hf \star J \cdot J,
\qquad
C_H= L\cdot L.
\label{CasimirSL}
\ee
Nowadays there are three known ways to get the constraints \eqref{Casconstr},
all of them leading to the same result \cite{Conrady:2010vx}.
Note that in the Lorentzian case it is possible also to take the subgroup $H=SU(1,1)$
which would describe a triangulation with tetrahedra having spacelike normals \cite{Alexandrov:2005ar}.
The EPRL model has been extended to this situation in \cite{Conrady:2010kc} and the resulting constraints
have been shown to have the same form \eqref{Casconstr}.

A concrete solution of the conditions \eqref{Casconstr} on the Casimir operators depends on
the signature and the value of the Immirzi parameter $\im$, which we assume to be positive in the following.
The general feature is however that these conditions do not have any solutions in terms of unitary representations.
Therefore, the usual strategy \cite{Engle:2007wy} is to adjust the values of
the Casimir operators by linear and constant terms in representation labels in such a way that solutions do exist.
The adjustment is then interpreted to be due to the ordering ambiguity at quantum level.
As we will see however, this does not fix the ambiguity in a unique way and we will argue in particular
that the standard EPRL solution should be slightly modified. In other words, both constraints \eqref{Casconstr}
are solved usually only approximately for large $j$.

At this point it is useful to note also that the generators of $G$ in any representation $\lambda$
satisfy the relation \eqref{relgen} \cite{GMS}. Moreover, one can check that the coefficient $\beta_{\lambda,j}$
can be expressed through Casimirs as follows
\be
\beta_{\lambda,j}=\frac{C^{(2)}_G(\lambda)}{2C_H(j)}.
\label{formbeta}
\ee
Then the second condition in \eqref{Casconstr} implies that $\beta_{\lambda,j}=\im$, exactly or approximately
according to how the constraint was solved. Assuming that the solution is exact,
one immediately gets the relation \eqref{projcon} where
\be
A^a=\hf\,{\eps^a}_{bc}\omega^{bc}-\im \omega^{0a}
\ee
is the AB connection. Thus, one does have a possibility to extract the LQG connection from the EPRL constraints.
But this requires the {\it exact} solution of one of the constraints. Since \eqref{formbeta} is just a fact
of representation theory and the relation \eqref{projcon} is linear in generators, there does not exist any ordering
ambiguity which could be used to relax this condition. On the other hand, as we mentioned, the EPRL model suggests
only an approximate solution. Below we show how this can be cured by appropriately modifying the resulting relations
between representation labels.

\subsubsection{Lorentzian theory}

For the Lorentz group the principle series irreducible representations are labeled by two numbers $\lambda=(n,\rho)$
with $n\in\Nint/2$, $\rho\in\bR$.
In our normalization\footnote{Our normalization is related to the one of \cite{Engle:2007wy} as follows:
being expressed in terms of generators $C_{1,\rm there}=2C_{G,\rm here}^{(1)}$, $C_{2,\rm there}=2C_{G,\rm here}^{(2)}$
and $\rho_{\rm there}=2\rho_{\rm here}$, $n_{\rm there}=2n_{\rm here}$.}
the Casimir operators read
\be
C_G^{(1)}=n^2-\rho^2-1,
\qquad
C_G^{(2)}=2n\rho,
\qquad
C_{SU(2)}= j(j+1)
\label{LCas}
\ee
and reducing to the $SU(2)$ subgroup one finds only representations with $j- n\in \Nint$.
Plugging the Casimirs \eqref{LCas} into \eqref{Casconstr}, one indeed finds that there are no solutions with
half-integers $j$ and $n$. The proposal of \cite{Engle:2007wy} is to take
\be
\rho=\im n, \qquad j=n
\label{solL}
\ee
which solves \eqref{Casconstr} up to linear terms in $j$. However, this gives $\beta_{\lambda,j}=\frac{\im j}{j+1}$
which does not allow to get the AB connection. Therefore we need a different solution. It can be fixed uniquely if one requires
that $\beta_{\lambda,j}=\im$ and $j$ is given by the lowest weight representation. Thus, we propose to replace
\eqref{solL} by
\be
\rho=\im(n+1), \qquad j=n.
\ee

\subsubsection{Euclidean theory}

In this case the gauge group is $Spin(4)=SU(2)\times SU(2)$ so that the irreducible
representations are labeled by two half-integers $\lambda=(\jp,\jm)$. The Casimir operators are
\be
C_G^{(1)}=2\jp(\jp +1)+2\jm(\jm +1),
\qquad
C_G^{(2)}=2\jp(\jp +1)-2\jm(\jm +1),
\label{ECas}
\ee
and the representations of the diagonal subgroup satisfy $|\jp-\jm|\le j\le \jp +\jm$.
The solution of the EPRL constraints now splits into two classes according to whether $\im$ is larger or less than 1
and is given by \cite{Engle:2007wy}
\be
\jm=\left|\frac{\im-1}{\im+1}\right|\jp,
\qquad
j=\left\{
\begin{array}{c}
\jp +\jm \quad \im<1
\\
\jp -\jm \quad \im>1
\end{array}\right. ,
\ee
or can also be written as
\be
\jp=\hf\,(1+\im)j,
\qquad
\jm=\hf\, |1-\im|j.
\ee
Note that it implies that the Immirzi parameter $\im$ is quantized to be a rational number.

It is easy to check that whereas for $\im<1$ one has $\beta_{\lambda,j}=\im$, for $\im>1$ one finds
$\beta_{\lambda,j}=\frac{\im j+1}{j+1}$. Thus, in the latter case, if one wants to get the AB connection, the solution
must be modified. At the same time, it is natural to keep the property that it selects the lowest
weight representation of $SU(2)$. This fixes it to be given by
\be
\im>1: \quad \jm=\frac{\im-1}{\im+1}(\jp+1), \qquad {\rm or} \qquad
\jp=\hf\,(\im+1)(j+1)-1,
\quad
\jm=\hf\,(\im-1)(j+1).
\ee

\bigskip

The same exercise can be repeated for the subgroup $H=SU(1,1)$ permitting triangulations with timelike surfaces
\cite{Alexandrov:2005ar,Conrady:2010kc}. We recapitulate all results in the following table which
gives the restrictions on representations leading to the AB connection for all possible choices of
groups:\footnote{In the case $H=SU(1,1)$ the label $j$ of the discrete series differs from $j$ in \cite{Conrady:2010kc}.
Its range is from $0$ to $n-1$ so that the constraints select the highest weight representation. For the continuous series
$j=-1/2+i s$. With these conventions in all cases $C_H=j(j+1)$. }

\vspace{0.3cm}\hspace{-2.0cm}
\begin{tabular}{|c|c|c|c|c|c|}
\hline
gauge group $G$ &
\multicolumn{2}{c|}{$Spin(4)$} &
\multicolumn{3}{c|}{$SL(2,\Cmat$)}
\\ \hline
\begin{tabular}{c}
subgroup $H$ \\ irreps, $\im$
\end{tabular} &
\begin{tabular}{c}
$SU(2)$ \\ $\im<1$
\end{tabular} &
\begin{tabular}{c}
$SU(2)$ \\ $\im>1$
\end{tabular} &
$SU(2)$ &
\begin{tabular}{c}
$SU(1,1)$ \\ discrete series
\end{tabular} &
\begin{tabular}{c}
$SU(1,1)$ \\ continuous series
\end{tabular}
\\ \hline
constr. on $\lambda$ &
$\jm=\frac{1-\im}{1+\im}\jp$ &
$\jm=\frac{\im-1}{\im+1}(\jp+1)$ &
$\rho=\im(n+1)$ & $\rho=\im(n-1)$ & $\rho=-n/\im$    \rule{0pt}{0.5cm}
\\ \hline
constr. on $j$ & $j=\jp+\jm$ & $j=\jp-\jm$ & $j=n$ & $j=n-1$ & $s^2+1/4=\rho^2$ \rule{0pt}{0.5cm}
\\ \hline
\end{tabular}

\subsubsection{Casimir operators}

As we mentioned above, all our restrictions on representations solve the initial constraints
\eqref{Casconstr}, where the Casimir operators are given by the usual expressions \eqref{LCas} and \eqref{ECas},
only up to linear and constant terms in the representation labels. This is attributed to the ordering
ambiguity in the Casimir operators which can be used to cancel the remaining terms \cite{Engle:2007wy}.
However, this is a non-trivial fact that such a cancelation is indeed possible given that
the adjusted operators should be independent of the Immirzi parameter $\im$.
Essentially, this is what allows to fix the ordering ambiguity in a unique way.
Here we give expressions for the Casimir operators, corresponding to the different cases considered above,
such that our restrictions on the representations solve the constraints \eqref{Casconstr} {\it exactly}.
\begin{itemize}
\item Lorentzian theory:
\be
C_G^{(1)}=(n+1)^2-\rho^2,
\qquad
C_G^{(2)}=2(n+1)\rho,
\qquad
C_{SU(2)}= (j+1)^2.
\label{LCasmod}
\ee
\item Euclidean theory, $\im<1$:
\be
C_G^{(1)}=2(\jp)^2+2(\jm)^2,
\qquad
C_G^{(2)}=2(\jp)^2-2(\jm)^2,
\qquad
C_{SU(2)}= j^2.
\label{ECas1mod}
\ee
\item Euclidean theory, $\im>1$:
\be
C_G^{(1)}=2(\jp+1)^2+2(\jm)^2,
\qquad
C_G^{(2)}=2(\jp+1)^2-2(\jm)^2,
\qquad
C_{SU(2)}= (j+1)^2.
\label{ECas2mod}
\ee
\end{itemize}
The cases with the subgroup $SU(1,1)$ can also be considered and
lead to similar expressions.

It is interesting to note two observations. First, the area spectrum following from the adjusted $SU(2)$ Casimir operator
is always equally spaced. Second, the ordering seems to change discontinuously at $\im=1$ in the Euclidean theory.
Whether these observations have a significant meaning deserves further investigation.

\subsection{Projected holonomies and projected connections}
\label{subsec_proj}

Although we showed how the AB connection can be obtained from the spin-connection
by using the constraints of the EPRL model on representations, so far this is just
a curious mathematical observation which does not allow to conclude that the EPRL states
are functionals of $A^a$. The problem is that in projected spin networks
the projectors $\pr{j}$ are inserted only at vertices. As a result, one finds
only combinations \eqref{projW} given by projected holonomies, whereas we need
``holonomies of a projected connection".

However, if one takes seriously the above results, one could ask whether there is a natural way
to get such objects from the state space of the EPRL model. It turns out it does exist
and can be found in the work \cite{Alexandrov:2002xc} where the covariant projection, crucial
for the definition of projected spin networks, has been introduced for the first time.
In that work it was suggested to consider fully projected holonomies obtained by inserting the projectors $\pr{j}$
along the whole integration path
\be
\Umat^{(\lambda,j)}_{\alpha}= \lim\limits_{N \rightarrow \infty}
{\cal P}\left\{ \prod\limits_{n=1}^{N}
\pr{j} U_{\alpha_n}^{(\lambda)} \pr{j} \right\},
\label{WLp}
\ee
where one takes the limit of infinitely fine partition $\alpha=\bigcup_{n=1}^N \alpha_n$.
It is easy to see that the projectors can be exponentiated and the resulting object is equivalent to
the holonomy of the projected connection \cite{Alexandrov:2002xc}
\be
\Umat^{(\lambda,j)}_{\alpha}[\omega]=
\iota_{\lambda}\left( U_{\alpha}^{(j)}[A]\right),
\label{WL-W2}
\ee
where $\iota_{\lambda}$ denotes the embedding of an operator
in a representation of $H$ into representation $\lambda$ of $G$.
Coupling these holonomies by intertwiners $\CI_v$, one recovers the usual spin networks of LQG.

In fact, one can get even a stronger result. So far we considered a gauge fixed version of our story
where, in the language of spin foams, all normals to tetrahedra (dual to the vertices of the boundary spin network)
were time-directed, $x_v^I=(1,0,0,0)$. What does change if one relaxes this condition?
First, a fixed unit-vector $x^I$ defines a subgroup $H_{x}\subset G$
which is the isotropy subgroup of this vector in 4d. In particular, in the Lorentzian case
it can be $SU(2)$ if $x^I$ is timelike or $SU(1,1)$ if $x^I$ is spacelike.
The relation \eqref{projcon} is then easily generalized to \cite{Alexandrov:2002br}
\be
\Ppr{j}{x}\(\omega_i^{IJ} J_{IJ}^{(\lambda)}\) \Ppr{j}{x}=A_i^a L_{x,a}^{(j)}\, ,
\label{projconx}
\ee
where the index $x$ indicates that these objects are defined with respect to the rotated subgroup $H_x$.
This allows immediately to extend the above results to the case where all normals $x_v$ are equal.
The general case can be obtained by applying a $G$-transformation and gives rise to holonomies of
the covariant generalization $\SSA^{IJ}$ of the AB connection introduced in \cite{Alexandrov:2001wt}
and used in \cite{Alexandrov:2002br} to formulate LQG in a Lorentz covariant form.
It is given by
\be
\SSA_i^{IJ}=
I_{(Q)KL}^{IJ}(1-{\im}\star) \omega_i^{KL}
+2(1+\im\star)x^{[J}\p_i x^{I]},
\label{conSU2}
\ee
where
\be
I_{(Q)}^{IJ,KL}(x)=
\eta^{I[K}\eta^{L]J}-2\sigma\, x^{[J}\eta^{I][K} x^{L]}
\label{projxxx}
\ee
is the projector on the Lie subalgebra of $H_{x}$, and has only nine independent components coinciding
for constant $x^I$ with $A_i^a$. Its appearance is guaranteed by the Lorentz invariance
of projected spin networks, but we also give a direct proof in appendix \ref{apend}.

Note that the projection \eqref{WLp} is incompatible with how the gauge invariance is
incorporated into the EPRL model. Usually, it is represented as the closure constraint
requiring that at each vertex the generators associated to the adjacent edges sum to zero,
which is equivalent to the $G$-invariance of all intertwiners.
The invariance is achieved by integrating over the normals $x_v$. However, to be able to define
the fully projected holonomy \eqref{WLp} along an edge, one should prescribe how the normal $x$ changes
along this edge since all projectors are defined with respect to this normal.
This function then appears in the resulting connection \eqref{conSU2}
and of course it must be smooth. This is possible only if $x$ remains an unintegrated free variable,
playing the role of an additional argument of the state functional.
This corresponds to a relaxed version of the closure constraint advocated in
\cite{Alexandrov:2007pq,Alexandrov:2008da}. It results in covariant intertwiners, but still invariant
spin network states.

Thus we conclude that by
\begin{itemize}
\item dropping integrals over the normals $x_v$,
\item making the projection \eqref{WLp} along the edges,
\end{itemize}
one can convert the state space of the EPRL model into the kinematical Hilbert
space of (the Lorentz covariant version of) LQG. Whereas the physical interpretation of the first step is clear
(it corresponds just to the gauge fixing in the path integral), the second step is somewhat mysterious.
In \cite{Alexandrov:2002xc} it was introduced to ensure that the resulting spin networks are eigenfunctions
of the area operator for surfaces which cross the graph at any point. Instead, the usual projected spin networks
are eigenfunctions only for those surfaces which are infinitely close to the vertices. This might be an
undesirable feature. From this point of view, the projection produces states with a more transparent
geometric interpretation.
On the other hand, the projection in \eqref{WLp} can be seen as the insertion of infinitely many bi-valent
vertices in the original edge of the graph.
This hints that it might be related to some kind of continuum limit of the model, although
it is not clear why the limit should affect the states in such peculiar way.

\section{Simplicity constraints revisited}
\label{sec_simple}

Although the results obtained in the framework of the EPRL and FK spin foam models are very promising
\cite{Conrady:2008mk,Barrett:2009gg,Bianchi:2009ri}, we would like now to reconsider their derivation.
There are several ways to get these models, but all of them rely on the common strategy:
{\it ``first quantize and then constrain"}. In our context this strategy is applied
to Plebanski formulation of general relativity in the presence of the Immirzi parameter $\im$.
It implies that, first, one quantizes the BF part of the theory and then imposes the simplicity constraints
\be
 B^{IJ}\wedge B^{KL}= \sigma{\cal V} \, \eps^{IJKL},
\label{simplicityconditions1}
\ee
with $\CV=\frac{1}{4!}\,\tr (B\wedge B)$ being the 4-dimensional volume form,
restricting the bi-vectors to be given by a tetrad, $B=\star(e\wedge e)$, already at the quantum level.
The last step requires a map from the classical constraints to their quantum version which is achieved
by promoting the bi-vectors $B$ to quantum operators. Following the above strategy, all SF models
use the map provided by the first step, the quantization of BF theory, which implies that
the bi-vectors can be identified with a particular combination of generators of the gauge algebra
determined by the Immirzi parameter
\be
B+\frac{1}{\im}\star B\ \mapsto\ J
\quad\mathop{\Leftrightarrow}\limits^{\im^2\ne \sigma}\quad
B\ \mapsto\ \frac{\im^2}{\im^2-\sigma}\(J-\frac{1}{\im}\star J\).
\label{newquantB}
\ee
Then different models propose different ways to impose the constraints \eqref{simplicityconditions1}.
For example, in the EPRL approach they are split into the diagonal and cross simplicity and
treated as being of first and second class, respectively. In the FK model instead
one requires the simplicity of expectation values of the quantized bi-vectors between certain coherent states.

Before discussing the weak points of this procedure, we propose as a warming-up to consider
a simple quantum mechanical model.
Despite its simplicity, it is able to capture the basic features of the SF quantization
of 4d general relativity and clearly identifies its loopholes.

\subsection{A simple example}
\label{subsec_example}

Let us consider a system described by the following action:
\be
S=\int dt \[ p_1 \dot q_1 +p_2 \dot q_2 -\hft p_1^2-\cos q_2+\lambda (p_2-\gamma p_1) \].
\label{ex-act}
\ee
Here the coordinates $q_1$ and $q_2$ are supposed to be compact, so that we consider them as
living in the interval $[0,2\pi)$, and $\gamma$ is a numerical parameter.

The canonical analysis of this system is elementary.
The momenta conjugate to $q_1$ and $q_2$ are $p_1$ and $p_2$, respectively, so that
the only non-vanishing Poisson brackets are
\be
\{ q_1,p_1\}=1,
\qquad
\{ q_2,p_2\}=1.
\label{Poisson}
\ee
The variable $\lambda$ is the Lagrange multiplier for the primary constraint
\be
\label{primconst}
\phi=p_2-\gamma p_1=0.
\ee
Commuting this constraint with the Hamiltonian
\be
H=\hft \,p_1^2+\cos q_2-\lambda\phi,
\ee
one finds the secondary constraint
\be
\psi=\sin q_2=0.
\label{secconst}
\ee
The latter constraint has two possible solutions
\be
q_2=0 \quad {\rm or} \quad q_2=\pi.
\label{twosol}
\ee
Since the two constraints, $\phi$ and $\psi$, do not commute,
they are of second class. A way to take this into account is to
construct the Dirac bracket. It is easy to find that the only
non-vanishing Dirac brackets between the original canonical variables are
\be
\{ q_1,p_1\}_D=1,
\qquad
\{ q_1,p_2\}_D=\gamma.
\label{Dirac}
\ee
The second bracket here is actually a consequence of the first one provided
one uses $p_2=\gamma p_1$. The Hamiltonian is given (up to a constant) by
\be
H=\hft\, p_1^2.
\ee
Thus, it is clear that the system reduces to the very simple system
describing one free particle on a circle.

The quantization of this system is also trivial. All different quantization methods
such as reduced phase space quantization, Dirac quantization and canonical path integral
lead to the same result that the $q_2$ degree of freedom is completely ``frozen"
and one has just a free particle with quantized momentum.
For example, following Dirac quantization, since $q_2$ is fixed, one represents the commutation
relations \eqref{Dirac} on functions of only $q_1$ as follows
\be
\hat q_1=q_1,
\qquad
\hat q_2=0,
\qquad
\hat p_1=-i\p_{q_1},
\qquad
\hat p_2=-i\gamma\p_{q_1}.
\label{repr_D}
\ee
The Hilbert space consists from periodic functions and its basis is provided by
\be
\Psi_{j_1}(q_1)=e^{ij_1q_1}.
\label{resst_D}
\ee
The Hamiltonian is represented simply as
\be
\hat H=-\p_{q_1}^2.
\label{HamDir}
\ee
It is easy to see that this quantization also agrees with the path integral method, which starts from
the phase space path integral and gives for correlation functions the following
result:\footnote{Here we neglected the contribution from the second
solution in \eqref{twosol} which has essentially the same form.}
\beq
<\CO>&=&\int dq_1 dq_2 dp_1 dp_2 \,|\det\{\phi,\psi\}|\, \delta(\phi)\delta(\psi)\,
e^{i\int dt\( p_1 \dot q_1 +p_2 \dot q_2 -\hf p_1^2-\cos q_2 \)} \CO(q_1,p_1,q_2,p_2)
\nonumber \\
&=& \int dq_1 dp_1 \,
e^{i\int dt\( p_1 \dot q_1 -\hf p_1^2 \)} \CO(q_1,p_1,0,\gamma p_1).
\eeq

Now we would like to consider what one obtains if one follows the spin foam strategy
to quantization. This question is reasonable because the model \eqref{ex-act}
can be viewed as a simplified version of Riemannian general relativity
with a finite Immirzi parameter. Indeed, $q_1$ and $q_2$ are analogous to the right and
left parts of the $SO(4)$ spin-connection under chiral decomposition, $p_1$ and $p_2$ correspond to
the chiral parts of the $B$ field, $\phi$ is similar to the diagonal simplicity
constraint, and $\gamma$ plays the role of the Immirzi parameter
(or rather of its combination $\frac{\gamma+1}{\gamma-1}$).

Thus, proceeding as in the new SF models, one should first drop the constraints generated by $\lambda$ and
quantize the remaining action. In the coordinate representation a basis in the Hilbert space is then
given by
\be
\Psi_{j_1,j_2}(q_1,q_2)=e^{ij_1q_1+ij_2q_2},
\label{enstates}
\ee
where due to the compactness of $q_1,q_2$ the labels $j_1,j_2$ are integer,
and the canonical variables are represented by operators satisfying the Poisson commutation relations \eqref{Poisson},
not the Dirac algebra:
\be
\hat q_1=q_1,
\qquad
\hat q_2=q_2,
\qquad
\hat p_1=-i\p_{q_1},
\qquad
\hat p_2=-i\p_{q_2}.
\label{repr_P}
\ee
At the second step, one imposes the constraint $\phi$ \eqref{primconst} requiring that
the states \eqref{enstates} should satisfy
\be
\hat\phi \,\Psi_{j_1,j_2}=(\hat p_2-\gamma\hat p_1)\Psi_{j_1,j_2}=0.
\ee
One immediately concludes that this gives a condition on the basis labels
\be
j_2=\gamma j_1,
\label{condj}
\ee
so that the physical states are spanned by
\be
\Psi_{j_1}(q_1,q_2)=e^{ij_1(q_1+\gamma q_2)}.
\label{resst_P}
\ee
These states can be viewed as analogues of the LQG spin networks with $q_1+\gamma q_2$ being similar to
the AB connection. Moreover, since $j_1,j_2$ are integer, the condition \eqref{condj} implies that
the parameter $\gamma$ should be a rational number, precisely as it happens in the Riemannian SF models for the Immirzi parameter.

This fact clearly shows that the quantization {\it \`a la} spin foam is not equivalent
to all other quantization methods. The difference can be noticed already in the form of the physical
states since the functions \eqref{resst_D} and \eqref{resst_P} depend on different classical variables.
Although it is tempting to identify them since the difference vanishes due to the constraint $\psi$,
it is nevertheless important because it affects the correlation functions involving $q_2$.

A more drastic discrepancy is that the parameter $\gamma$ is quantized in the SF approach and
does not have any restrictions in the usual quantization. This problem cannot be avoided by any tricks
and shows that the two quantizations are indeed inequivalent. Taking into account
the classical analysis and the fact that the first approach represents actually
a result of several possible methods, which all follow the standard quantization rules,
it is clear that it is the first quantization that is more favorable and
the quantization of $\gamma$ does not seem to have any physical reason behind itself.

In fact, it is easy to trace out where the SF approach diverges from the standard one:
it takes too seriously the symplectic structure given by the Poisson brackets.
On the other hand, it is the Dirac bracket that describes the symplectic structure
which has a physical relevance. In particular, in the presented example, the Poisson
structure tells us that $p_2$ is the momentum conjugate to $q_2$, whereas
in fact it is conjugate to $q_1$.

This ignorance of the right symplectic structure has serious consequences.
Let us take a look not only at the physical Hilbert space, but also at the Hamiltonian acting on it.
In the approach {\it \`a la} spin foam it reads
\be
\hat H=-\p_{q_1}^2+\cos q_2.
\label{wrongH}
\ee
But this operator is simply not defined on the subspace spanned by linear combinations of \eqref{resst_P}!
The problem is caused by the second term involving $q_2$.
It is impossible to ignore this term by requiring in addition that
$\hat \psi$ vanishes on the physical states since it would be in contradiction with
the commutation relations \eqref{repr_P}.

Moreover, let us assume that one succeeded somehow to define the Hamiltonian operator on the physical subspace.
But then it would not have eigenstates there!
Indeed, from $\hat\phi \Psi=0$, one gets
\be
( [\hat H,\hat\phi] + \hat\phi \hat H )\Psi=(\hat\psi + \hat\phi \hat H )\Psi=0,
\label{eigenH}
\ee
where we used the definition of the secondary constraint.
Assuming that $\Psi$ is an eigenstate of $\hat H$ would lead to
the condition $\hat \psi \Psi = 0$.
But as we mentioned above, this is not consistent with \eqref{repr_P}.

Thus, we have to conclude that the SF strategy applied to the simple system \eqref{ex-act}
leads to a quantization which is not simply different from the usual one,
but intrinsically inconsistent.
This inconsistency is just a manifestation of the fact that the rules of the Dirac quantization, and in particular
the necessity to use the symplectic structure modified by the second class constraints, cannot
be avoided. In our opinion, this is the only correct way to proceed leading to a consistent quantum theory.

\subsection{Simplicity constraints in spin foam models}
\label{subsec_SF}

These conclusions can be immediately translated to most of the spin foam models
since they all quantize the symplectic structure provided by BF theory
which ignores constraints of general relativity. Indeed, in that theory, or more precisely in its $\im$-deformed
version, the combination $B+\frac{1}{\im} \star B$ is the variable conjugate to the connection.
This fact is the underlying reason for the identification \eqref{newquantB} crucial
for the implementation of the simplicity constraints \eqref{simplicityconditions1} at the quantum level.
On the other hand, Plebanski formulation,
supposed to be the starting point of the SF approach, contains second class constraints
leading to Dirac brackets and modifying the symplectic structure
\cite{Buffenoir:2004vx,Alexandrov:2006wt}.
The second class constraints come in pairs: a half of them are the simplicity constraints
\eqref{simplicityconditions1} and the second half appears as secondary constraints obtained by commuting
\eqref{simplicityconditions1} with the Hamiltonian, precisely as in the example above.
In particular, the secondary constraints include
\be
\Psi^{ij}=\eps^{mn(i}\eps^{j)kl}\eps_{IJKL} B_{0n}^{IJ}D_m B_{kl}^{KL},
\label{secconstr}
\ee
where $D_n$ is a covariant derivative defined by $\omega^{IJ}$ and $(\cdot \cdot )$ denotes symmetrization.
(The full list of constraints can be found in \cite{Buffenoir:2004vx}.)
The constraint \eqref{secconstr} is in fact equivalent to a similar constraint
of the usual Hilbert--Palatini formulation with the Immirzi parameter \cite{Alexandrov:2000jw}, so do the symplectic
structures of the two formulations \cite{Alexandrov:2006wt}.
(See also \cite{Engle:2009ba} for a relation at the path integral level.)
The modification of the symplectic structure by second class constraints has important implications for us
because it implies a completely different identification
between the $B$ field and the generators of the gauge group \cite{Alexandrov:2007pq}.
In particular, the new identification ensures that the simplicity constraints are satisfied automatically,
as it should be for any second class constraint quantized using Dirac brackets.
However, this fact as well as the necessity to take into account the secondary constraints have
been overlooked so far in the SF approach.

Moreover, the ignorance of the secondary constraints leads to another common confusion.
The simplicity constraints in their covariant form \eqref{simplicityconditions1} are certainly enough to get
general relativity from BF theory.
However, to impose them at the quantum level, one needs to understand
which of them are first and which are second class.
Usually in the SF approach, their class is determined by the
commutation algebra of these constraints after quantization according to the map \eqref{newquantB}.
As a result, one finds that their algebra contains a center, the diagonal simplicity, which
is interpreted as first class,
whereas the remaining cross simplicity constraints are considered as second class \cite{Engle:2007wy}.
But such analysis assumes that the simplicity constraints exhaust all set of constraints of the theory.
But this is not true! As we saw, the classical theory contains also secondary constraints imposing
some conditions on the connection variable and similar to the constraint $\psi$ \eqref{secconst}
in the above example.
Taking them into account, all simplicity are always of second class and the proper way to incorporate them is through
the symplectic structure given by the Dirac brackets.

These considerations show the incompatibility of the SF strategy with the Dirac approach.
Besides, we see that both EPRL and FK models can be viewed as (very non-trivial) generalizations
of the example considered in the previous subsection:
it captures the $U(1)\times U(1)$ sector of the theory constrained by diagonal simplicity.
The non-diagonal degrees of freedom are missed, but their presence cannot solve problems caused by
the incorrect treatment of the diagonal simplicity and by the use of a wrong symplectic structure.
Therefore, it is natural to expect that these models
are supplied with inconsistencies of the kind presented in the previous subsection.
In particular, the quantization of $\im$ in the Riemannian models is one of their manifestations.
One may think that this is a default of only Riemanninan models
since there is no quantization condition on $\im$ in the Lorentzian case. But it is clear that this is
just a particular consequence of the generic strategy employed by SF models and it is the strategy
that is problematic, rather than the choice of the signature, which can be viewed simply as a condition of
the compactness of some variables.

Other types of inconsistencies are expected to arise already at the dynamical level
as follows from the discussion around \eqref{wrongH}. More precisely, in the canonical picture corresponding
to the new SF models, the existence of secondary constraints \eqref{secconstr} implies
that the Hamiltonian operator would not be well defined on the subspace defined by the quantum
simplicity constraints \eqref{Casconstr}.
In particular, its kernel, which should encode the physical state space of quantum gravity, cannot belong to this subspace.
These features however do not prevent us from considering matrix elements of this Hamiltonian operator
and defining a vertex amplitude. But given the above properties and the fact
that it would differ from the one obtained following the Dirac rules
(cf. Hamiltonians in \eqref{HamDir} and \eqref{wrongH}), the physical relevance of such vertex is quite suspicious.

Thus, all new SF models appear to suffer from applying the strategy which is inconsistent
with the rules of quantum mechanics. There is however one special case which is potentially free from
the above problems. This is the FK model without the Immirzi parameter ($\im=\infty$) \cite{Freidel:2007py}.
This model is constructed as a path integral quantization of the discretized Plebanski theory.
Although the implementation of the simplicity constraints in this model is also
based on the map \eqref{newquantB}, in the path integral the $B$ field appears only through its expectation values in
coherent states \cite{Livine:2007vk}. It is these expectation values that are required to satisfy
the simplicity and therefore some components of the $B$ field turn out to be projected out. For $\im=\infty$
at each tetrahedron (dual to a vertex of the boundary spin network)
one obtains the following effective quantization rule
\be
B\ \mapsto \ I_{(P)}(x_v)\cdot J,
\label{qmap}
\ee
where $I_{(P)}^{IJ,KL}(x)=2\sigma\,x^{[J}\eta^{I][K} x^{L]}$ is the projector on the orthogonal completion
of the isotropic subalgebra of the vector $x^I$.
This identification is consistent with the symplectic structure
of Plebanski formulation written in terms of
a shifted connection \cite{Alexandrov:2001wt,Alexandrov:2006wt}.
Thus, this model seems to be able to provide the right kinematics.

In this respect and for this particular parameter $\im$ the FK model differs crucially from the EPRL model.
Taking into account that for $\im=\infty$ the EPRL model reduces to the Barrett-Carne model \cite{Barrett:1997gw},
which is known to be incapable to describe quantum gravity, it may be not surprising that
the FK model appears to be more favorable.
However, as was argued in \cite{Alexandrov:2008da}, it still ignores the presence
of the secondary second class constraints which may affect the resulting vertex amplitude.
It seems that to get a complete model one should really abandon the usual SF strategy and,
starting from the very beginning, to find a way leading to the SF representation.

\section{Conclusions}
\label{sec_concl}

In this paper we arrived at two opposite, but not contradictory conclusions.
On one hand, we found that the formal coincidence of the state space of the EPRL spin foam model
with the kinematical Hilbert space of LQG can be deepen in such a way that two corresponding states
are represented by the same functionals of the same configuration variable,
which is the AB connection or its covariant generalization. This however requires three things:
\begin{itemize}
\item  an adjustment of the restrictions on representations, which thereby fixes the ordering ambiguity
of the EPRL approach;
\item dropping the integral over the normals living at vertices;
\item  a certain projection along all edges of spin networks, which converts them to eigenstates
of area operators associated with {\it any} 2d surface.
\end{itemize}
The last point shows that the relation between SF and LQG states is not so direct.
Although the projection is a one-to-one map (if one does not allow bi-valent vertices on the SF side),
its physical meaning is obscure.
Nevertheless, this result can be considered as an additional indication in favor of the
potential convergence of the spin foam and loop quantizations.

On the other hand, our second conclusion is that the current spin foam strategy to quantization in four dimensions,
summarized as {\it ``first quantize and then constrain"},
is not viable. In our opinion, it contradicts to the quantization rules for systems with second class constraints
and leads to various inconsistencies at the quantum level.
In particular, we believe that the new spin foam models \cite{Freidel:2007py,Engle:2007wy} do not provide
a proper quantization of general relativity.
This conclusion has however one exception which is the FK model for $\im=\infty$, to which our analysis cannot be applied.
At present it can be considered as the best candidate for the correct spin foam model, which avoids
the basic problems of other models and gives the right kinematics. But even in this case one expects that
the dynamics is not captured in a satisfactory way \cite{Alexandrov:2008da}.
Thus, the quest for the right model is to be continued.

\section*{Acknowledgements}

The author is grateful to F. Conrady, L. Freidel and Ph. Roche for valuable discussions
and to Perimeter Institute for Theoretical Physics for the kind hospitality and the financial support.
This research is supported by contract ANR-09-BLAN-0041 and in part by Perimeter Institute.

\appendix

\section{Covariant AB connection}
\label{apend}

The aim of this appendix is to show how the projection \eqref{WLp} generalized to the normal $x^I$
varying along the integration path gives rise to the covariant AB connection \eqref{conSU2}.
Thus, we should consider
\be
\Umat^{(\lambda,j)}_{\alpha}= \lim\limits_{N \rightarrow \infty}
{\cal P}\left\{ \prod\limits_{n=1}^{N}
\Ppr{j}{x_{n+1}} U_{\alpha_n}^{(\lambda)} \Ppr{j}{x_n} \right\},
\label{WLpx}
\ee
where $x_n$ denote values of $x$ at the ends of segments $\alpha_n$.
For constant $x$, one can use \eqref{projconx} where the r.h.s. can be written as
\be
\Bigl(I_{(Q)KL}^{IJ}(1-{\im}\star) \omega_i^{KL} \Bigr) \(\Ppr{j}{x}J_{IJ}^{(\lambda)}\Ppr{j}{x}\).
\ee
The first factor gives the contribution to the projected connection and the second denotes
the projected generators. To get the result for varying $x$, one can take one factor
in the product \eqref{WLpx}, perform a gauge transformation which does not affect $x_n$ but
makes $x_{n+1}$ equal to $x_n$, and then use the result for constant $x$.
If $x_{n+1}-x_n=\delta x$, to the first order in $\delta x$ the inverse transformation is given by
\be
g_x=\exp\(2 x^{[J}\delta x^{I]} J_{IJ}\),
\ee
where one should take into account that $x^I$ is a unit vector.
Then one finds
\be
\begin{split}
\Ppr{j}{x_{n+1}} U_{\alpha_n}^{(\lambda)}[\omega] \Ppr{j}{x_n}
& = g^{(\lambda)}_x \Ppr{j}{x_{n}} U_{\alpha_n}^{(\lambda)}[\omega^{g^{-1}_x}] \Ppr{j}{x_n}
\\
& \approx  \(1+2 x^{[J}\delta x^{I]} J_{IJ}^{(\lambda)}\)\Ppr{j}{x_{n}}
\[1+ \delta x^i \( \omega^{IJ}_i-2 x^{[J}\p_i x^{I]} \)J_{IJ}^{(\lambda)}\] \Ppr{j}{x_{n}}
\\
& \approx \Ppr{j}{x_{n+1}} \[1+ \delta x^i \( I_{(Q)KL}^{IJ}(1-{\im}\star) \omega_i^{KL}
+2(1+\im\star)x^{[J}\p_i x^{I]} \)J_{IJ}^{(\lambda)}\] \Ppr{j}{x_n}
\\
& \approx \Ppr{j}{x_{n+1}} U_{\alpha_n}^{(\lambda)}[\SSA] \Ppr{j}{x_n}.
\end{split}
\ee
Thus, the effect of the projection is that one obtains a holonomy of an effective connection
which coincides with the covariant generalization of the AB connection from \cite{Alexandrov:2001wt}.

\providecommand{\href}[2]{#2}

\end{document}